\documentclass[12pt]{article}
\usepackage{graphicx}
\begin{document}
\baselineskip=20pt
\markboth
{\it S.K. Patra}
{\it S.K. Patra}
\mbox{ }

\begin{center}
{\Large
{\bf Ground-state properties and spins of the odd {\boldmath$Z=N+1$}
     nuclei {\boldmath$^{61}$}Ga{\boldmath$-$}{\boldmath$^{97}$}In} }
\\[4ex]
S.K. Patra, M. Del Estal, M. Centelles and X. Vi\~nas \\
{\it Departament d'Estructura i Constituents de la Mat\`eria,
     Facultat de F\'{\i}sica,
\\
     Universitat de Barcelona,
     Diagonal {\sl 647}, E-{\sl 08028} Barcelona, Spain}
\\[2ex]
\end{center}
\vspace*{2cm}

{\small
\centerline {\bf Abstract}
Binding energies, quadrupole deformation parameters, spins and
parities of the neutron-deficient odd $Z=N+1$ nuclei in the $A\sim 80$
region are calculated in the relativistic mean field approximation.
The ground-state and low-lying configurations of the recently observed
$^{77}$Y, $^{79}$Zr and $^{83}$Mo nuclei are analyzed. The calculated
results are compared with other theoretical predictions.

\bigskip
\bigskip
\bigskip

 }

\newpage

\section{Introduction}
\hspace*{\parindent}

Beyond $Z\approx 20$ the stability of nuclei requires additional
neutrons because of the Coulomb repulsion among protons and the most
stable nuclei are those with $N > Z$ \cite{bohr0}. However, a large
number of nuclei are possible whose $N$ and $Z$ numbers differ
considerably from this line of $\beta$--stability. The properties of
light systems near the limits of stability of proton/neutron-rich
nuclei have attracted considerable experimental and theoretical
attention  \cite{tani0,mohar0,janas0,patra0,lala2,lala3}. The
availability of radioactive beams in various laboratories will likely
provide many intriguing experimental information on the structure and
reactions of these nuclei. The discovery of new isotopes \cite{mohar0}
has opened a new path of nucleosynthesis by rapid proton capture
\cite{patra0}. Similarly, the discovery of new neutron-rich nuclei
near the drip line is important to understand the rapid neutron
capture process in accreting stellar systems \cite{buer0}.

In the region of $A\sim 80$, nuclei with nearly equal number of
protons and neutrons are of fundamental interest and can now be
studied using radioactive beams \cite{blank0}. The structural
properties of these nuclei are strongly determined by deformed shell
gaps in the nuclear single-particle potential \cite{naza0}. The
deformation properties of these nuclei change dramatically by addition
or removal of one or two nucleons \cite{lister0,gelletly0}. The
nucleon numbers ($N$ or $Z$) 36 and 38 have been identified with
highly deformed oblate \cite{chandler0} and prolate
\cite{lister0,lister1} shell gaps, respectively. Recently, the very
neutron-deficient $Z=N+1$ ($T_z= -1/2$) nuclei $^{77}$Y, $^{79}$Zr and
$^{83}$Mo  have been observed \cite{janas0}. The deformation
properties of these nuclei and the energies of the last occupied
single-particle state of the odd proton are very crucial from
stability and astrophysical points of view.

Wallace and Woosley \cite{wallace0} have conjectured a rapid proton
capture process in accreting matter that provides a way for
synthesizing very neutron-deficient nuclei close to the proton drip
line in the $A\approx 60$--80 region \cite{mathews0}. In this case the
asymmetry energy is relatively unimportant because of the near
equality of $Z$ and $N$. The existence of these new highly
neutron-deficient isotopes stems from a delicate balance between the
attractive nuclear force and the repulsive electrostatic force in
atomic nuclei. On average, the nuclear force is attractive between a
proton and a neutron and less attractive between two protons or two
neutrons. Thus there is a limit to the excess number of protons over
neutrons, or vice versa, one can have in a nucleus. This situation is
further aggravated by the electromagnetic Coulomb repulsion among
protons which strives to break the nucleus apart. The limits to the
number of protons/neutrons are known as the proton/neutron drip lines.
Due to the increasing importance from both the experimental and
theoretical sides of the mass region $A\sim 80$, it is worthwhile to
investigate the ground-state properties and spin of these nuclei,
which is the prime aim of this work.

The paper is organized as follows. Section 2 is devoted to some basic
points of the relativistic mean field (RMF) calculations. We present
our results obtained by various RMF parameter sets in Section 3.
Finally, the summary and concluding remarks are given in Section 4.

\section{Calculations}
\hspace*{\parindent}

We shall calculate the deformation properties and the single-particle
energies and spins of the last occupied proton states for odd $Z=N+1$
systems using an axially deformed relativistic mean field (RMF)
formalism \cite{ring0,patra1}. From the relativistic Lagrangian we get
the field equations for the nucleons and the mesons. These equations
are solved by expanding the upper and lower components of the Dirac
spinors and the boson fields in a deformed oscillator basis with an
initial deformation $\beta_0$. $N_{\rm F} = 12$ and $ N_{\rm B} =20$ oscillator
shells are used as the expansion basis for the fermion and boson
fields \cite{ring0}. The set of coupled equations is solved
numerically by a self-consistent iteration method. The centre-of-mass
motion is estimated by the usual harmonic oscillator formula. We
evaluate the one-proton separation energy ($S_{\rm p}$) from the
binding energies of the two neighbouring nuclei with $Z$ and $Z-1$
protons \cite{bohr0}:
\begin{equation}
 S_{\rm p}(N,Z)= B(N,Z) - B(N,Z-1) \,,
\label{eq1}\end{equation}
where $B(N,Z)$ is the binding energy for neutron number $N$ and proton
number $Z$. The quadrupole deformation parameter $\beta_2$ is
evaluated from the resulting quadrupole moment \cite{ring0}.

Our calculations will be performed with the NL1 \cite{rein0}, NL-SH
\cite{shar0}, TM1 \cite{toki0} and NL3 \cite{lala0} parameter sets.
The predictive power of these parametrizations is well known and some
examples can be found, e.g., in Ref.\ \cite{patra97} and references
quoted therein. It is to be noted that the RMF parameter sets are
determined by fitting nuclear matter properties, neutron-proton
asymmetry energies, root-mean-square radii and binding energies of
some spherical nuclei. Then, there is no further adjustment to be made
in the parameters of the Lagrangian. The NL1 set was preferred in
early calculations \cite{rein1}. However, it does not describe well
the neutron skin thickness of neutron-rich nuclei due to a very large
asymmetry energy, and predicts relatively large quadrupole
deformations near the neutron drip line \cite{shar0}. To cure these
deficiencies, data on neutron radii were included in the fit of the
parameters of the NL-SH interaction. An interesting feature of the TM1
parametrization \cite{toki0} is that in this set the sign of the
quartic scalar self-coupling is positive (contrary to NL1, NL-SH and
NL3). This could be achieved by introducing a quartic self-interaction
of the vector field in the effective force. In general the quality of
the results reproduced by TM1 is not superior to the standard
non-linear sets and it has not been much used in the literature. The
relatively new parameter set NL3 is considered to be very successful
and there is confidence that it can be used fruitfully for the
investigation of new regions of nuclear stability.

The calculation of odd-even and odd-odd nuclei in an axially deformed
basis is a tough task in the RMF model. To take care of the lone odd
nucleon one has to violate time-reversal symmetry in the mean field.
In the present study only the time-like components $V_{0}$, $b_{0}$
and $A_{0}$ of the $\omega$, $\rho$ and photon fields are retained.
The space components of these fields (which are odd under time
reversal and parity) are neglected. They are important in the
determination of properties like magnetic moments \cite{hofm1}, but
have a very small effect on bulk properties like binding energies or
deformations and can be neglected to a good approximation
\cite{lala1}. In our calculation of odd nuclei we employ the blocking
approximation, which restores the time-reversal symmetry. In this
approach one pair of conjugate states $\pm m$ is taken out of the
pairing scheme. The odd particle stays in one of these states and its
corresponding conjugate state remains empty. In general one has to
block in turn different states around the Fermi level to find the one
which gives the lowest energy configuration of the odd nucleus. In
odd-odd nuclei (which will be needed in our calculations of separation
energies) we have blocked both the odd proton and the odd neutron.

For known nuclei close to or not too far from the stability line, the
BCS approach provides a reasonably good description of the pairing
properties. However, in going to nuclei in the vicinity of the drip
lines the coupling to the continuum becomes important. It has been
shown that the self-consistent treatment of the BCS approximation
breaks down when coupling between bound states and states in the
continuum takes place \cite{doba0}. For most of the very
neutron-deficient nuclei of our study odd-even mass differences are
not measured and little is known about the precise effect of the
pairing interaction. It is expected that for odd-even nuclei the
effects of pairing are considerably decreased \cite{bohr0}. In the
present investigation we have chosen to use a BCS formalism with a
small constant pairing strength, namely $\triangle_{\rm
n}=\triangle_{\rm p} = 0.5$ MeV\@. This value of the gaps contributes
very little to the total binding (unless the pairing gap is varied
considerably our results remain unchanged). This type of prescription
has already been adopted in the past \cite{patra2}. 

Certainly, for properties like radii of halo nuclei that sensitively
depend on the spatial extension of the nucleon densities a more proper
treatment of the continuum could be crucial, e.g., by means of the
relativistic Hartree-plus-Bogoliubov (RHB) approach
\cite{meng0,meng1,lala1}. In this model the wave functions of the
occupied quasi-particle states have the correct asymptotic behaviour.
Results of RHB and RMF-BCS calculations have been compared in Ref.\
\cite{lala3} for neutron-rich nuclei in the deformed $N=28$ region.
The two models have been found to predict almost identical binding
energies and similar quadrupole deformations, though they differ
significantly in the calculated r.m.s.\ radii (they turn out to be
larger in the RMF-BCS model). A recent RHB study of deformed odd-$Z$
proton emitters in the $53 \leq Z \leq 69$ region using the NL3 set
has been published in Ref.\ \cite{lala1}. For the lightest isotopes
$^{107}$I, $^{108}$I and $^{109}$I reported in Table 1 of that work,
the odd valence proton occupies a $[422]{3/2}^{+}$ Nilsson orbital
(see below for notation) and the ground-state quadrupole deformations
are $\beta_2= 0.15$, 0.16 and 0.16, respectively. For comparison we
have performed the calculations with our model and find the same
$[422]{3/2}^{+}$ orbital for the three isotopes and deformations
$\beta_2= 0.17$, 0.18 and 0.19, respectively, in rather good agreement
with the more sophisticated RHB method.

\section{Results and discussion}
\hspace*{\parindent}

We now discuss the results of our RMF calculations for the
neutron-deficient nuclei $^{61}$Ga, $^{65}$As, $^{69}$Br, $^{73}$Rb,
$^{77}$Y, $^{81}$Nb, $^{85}$Tc, $^{89}$Rh, $^{93}$Ag and $^{97}$In,
i.e., the odd-proton, $T_z=-1/2$ nuclei in the interval $31 \leq Z
\leq 49$ with $Z=N+1$. For a given nucleus the solution with the
largest binding energy corresponds to the ground-state configuration
and the other solutions are the excited intrinsic states. In the
present calculations we find two or three different solutions for most
of the isotopes, each solution differing in the deformation from the
others. All the solutions are often close in energy with one another,
and sometimes they are nearly degenerate. In the case of finding
almost degeneracy there is some uncertainty in the determination of
the ground-state solution: a change in the inputs of the calculation
(e.g., the parameter $\hbar\omega = 41 A^{-1/3}$ MeV) may alter the
prediction for the ground-state shape. The low-lying excited solutions
can be interpreted as solutions with coexisting shapes. The shape
coexistence nature in the $A\sim 80$ region has been reported in
Refs.\ \cite{maha0,prah1}.

In Tables 1--3 we present our RMF results for the binding energy and
the quadrupole deformation parameter $\beta_2$. We also list the
single-particle energy $\epsilon_{\rm p}$ of the blocked state,
occupied by the odd proton, as well as its Nilsson state labelings $[N
n_{3} \Lambda] {\Omega^{\pi}}$ ($\Omega^{\pi}$ being the spin and
parity of the orbit, for spherical solutions we use spherical quantum
numbers). For these odd-mass nuclei the spin of the odd nucleon is the
resultant spin of the nucleus. In the tables we also display results
of microscopic-macroscopic (MM) mass models for comparison. The values
from the tabulation of Ref.\ \cite{moller0}, based on the finite-range
droplet model and folded Yukawa single-particle potential, will be
labelled by MMa. The microscopic-macroscopic calculations described in
Ref.\ \cite{janas0} (mass formula plus Strutinsky correction) will be
labelled by MMb.

The RMF calculations predict a moderate prolate and a moderate oblate
solution for the $^{61}$Ga, $^{65}$As and $^{69}$Br nuclei (Table 1).
The ground-state shape of $^{61}$Ga is prolate in all the four RMF
parameter sets, and the quadrupole deformation parameter $\beta_2\sim
0.22$ reproduces very well the value of the microscopic-macroscopic
MMa model. The ground-state spin is ${1/2}^{-}$ according to NL1 and
${3/2}^{-}$ according to NL-SH, TM1 and NL3. The last odd proton is
bound by $-1$, $-0.9$, $-1.6$ and $-0.9$ MeV, respectively. The MMa
model proposes a spin of ${1/2}^{-}$ for $^{61}$Ga, in agreement with
the NL1 prediction. It is to be noted that sometimes there are several
levels near the Fermi surface available to the odd proton. Then we
blocked those levels in turn and chose the solution which corresponds
to the maximum binding. However, we also noticed that two (or more,
typically in spherical configurations) different blocked solutions may
be very close in energy and deformation. In such cases it is
difficult to select the ground-state solution. For example, this
situation arises for the prolate shape of $^{61}$Ga with the NL3 set.
We find a binding energy of 511.27 MeV ($\beta_2= 0.225$) when we
block the $[312]{3/2}^{-}$ level, whereas the binding energy is 511.03
MeV ($\beta_2= 0.228$) when the level $[310]{1/2}^{-}$ is blocked. In
the tables we present the result which corresponds strictly to the
maximum binding. In the $^{65}$As nucleus the prolate and oblate
solutions have very similar energies. Excepting NL1 where the prolate
shape has a ${3/2}^{-}$ spin, the spin of both solutions is
${1/2}^{-}$. The ground state corresponds to the prolate shape, with a
deformation $\beta_2\sim 0.23$ as in the MMa model. The MMa spin is
${3/2}^{-}$, as with NL1. For the ground state of $^{69}$Br the
Nilsson orbital occupied by the odd proton is $[404]{9/2}^{+}$ in the
four relativistic sets, which agrees with the MMa \cite{moller0} and
MMb \cite{janas0} calculations. The RMF suggests an oblate $^{69}$Br
ground state with a deformation $\beta_2$ around $-0.29$, similarly to
the MM models.

For $^{73}$Rb, $^{77}$Y and $^{81}$Nb (Table~2) we find different
solutions that often have close binding energies. In detail, for the
$^{73}$Rb nucleus the most bound solution is oblate ($\beta_2\sim
-0.35$) and the proposed spin is ${7/2}^{+}$ in all the parameter
sets. The MM models, however, predict a prolate shape with
$\beta_2\sim 0.4$ and spin ${3/2}^{+}$ (MMa) or ${3/2}^{-}$ (MMb),
which agrees better with the RMF prolate solution. For $^{77}$Y we
find a large prolate deformation in the ground state, with the
exception of TM1 that predicts an oblate ground-state shape. Apart
from the case of NL1, we also find a spherical ${1/2}^{-}$ ($p_{1/2}$)
configuration that appears as an excited state, though for TM1 it
coexists with the oblate ground state. In TM1 the prolate solution
lies at an excitation energy of about 1 MeV\@. Ignoring this energy
difference, then $[422]{5/2}^{+}$ is the last proton orbit of $^{77}$Y
in all the parameter sets which is supported by the MM predictions.
Comparing the various solutions for $^{81}$Nb, we find that NL1 gives
a nearly spherical ${9/2}^{+}$ ground state, NL-SH and NL3 predict a
highly prolate ${1/2}^{+}$ ground state (like the MM models) and TM1
gives coexistent oblate and almost spherical ${9/2}^{+}$ shapes.
(Actually, configurations of spin ${7/2}^{+}$, ${5/2}^{+}$,
${3/2}^{+}$ and ${1/2}^{+}$ with nearly zero deformation are found
lying at energies very close to that of the ${9/2}^{+}$ configuration
and all them originate from the spherical $g_{9/2}$ shell.) If we take
into account that the RMF calculation has some uncertainty, or the
increase in binding after performing angular momentum projection
calculations (which is particularly sizeable for solutions with a
large deformation \cite{prah1}), and assuming the highly deformed
shape as the ground state, then the Nilsson orbit of the last occupied
proton is $[431]{1/2}^{+}$ in accordance with the
microscopic-macroscopic calculations. The single-particle energy of
the valence proton in the spherical state of $^{81}$Nb with the NL1
and TM1 sets is positive. In such a case the system is unstable
against proton emission and we have included these solutions only for
completeness. 

The RMF sets yield an oblate ground state for $^{85}$Tc (Table 3),
with $\beta_2= -0.22$ and a Nilsson orbit $[413]{5/2}^{+}$ for the
last occupied proton. There appears a nearly spherical ${3/2}^{+}$
configuration, which for NL1 is degenerate in energy with the oblate
shape. The MMa and RMF predictions do not agree, the latter being
closer to the MMb solution. The ground-state shape and the Nilsson
orbit are parameter dependent for the $^{89}$Rh nucleus (Table 3). NL1
suggests a $\beta_2= 0.16$ solution of spin ${5/2}^{+}$, similarly to
MMa, NL-SH points to a $\beta_2= -0.20$ shape of spin ${3/2}^{+}$, and
for TM1 and NL3 the prolate and oblate solutions nearly have the same
energy. The MMb model favours a spherical configuration. We find
low-lying prolate and oblate solutions for $^{93}$Ag. The oblate
solution is close to sphericity in the case of NL1 and TM1. The
ground-state corresponds to a [413]${7/2}^{+}$ orbit with a
$\beta_2\sim 0.14$ deformation. The RMF parameter sets predict nearly
spherical solutions for the $^{97}$In nucleus, due to approaching the
$Z=50$ magic number. For both $^{93}$Ag and $^{97}$In, the RMF and MMa
proposed ground states compare well.

Next we analyze the results for the recently observed $Z=N+1$, $T_z=
-1/2$ nuclei $^{77}$Y, $^{79}$Zr and $^{83}$Mo. The properties of
$^{77}$Y have already been presented earlier and those of $^{79}$Zr
and $^{83}$Mo are displayed in Table 4. In $^{79}$Zr and $^{83}$Mo the
last odd nucleon is a neutron, and the spin and parity are decided by
this last valence neutron. We found three solutions (prolate,
spherical and oblate) for $^{79}$Zr with all the parameter sets. NL1
and TM1 predict a spherical shape of spin ${1/2}^{-}$ for the ground
state, whereas NL-SH and NL3 favour a largely prolate ground state
($\beta_2\sim 0.5$) of spin ${5/2}^{+}$. The spin of the oblate
solutions ($\beta_2=-0.18$ for the four parameter sets) is
${9/2}^{+}$. The RMF ground-state solution for the $^{83}$Mo nucleus
prefers an oblate shape in the NL-SH, TM1 and NL3 sets with a spin of
${7/2}^{+}$. On the other hand, NL1 suggests an oblate (${7/2}^{+}$)
and almost spherical (${9/2}^{+}$) shape coexistence nature. Once
more, the properties of the ground state are in consonance with the
MMa predictions.

Amongst the odd-$Z$ nuclei studied here, only $^{61}$Ga, $^{65}$As,
$^{77}$Y and $^{89}$Rh have been observed in experiment (see Janas et
al.\ \cite{janas0} and references [6]--[12] quoted therein). The
experimental evidence suggests that $^{69}$Br, $^{73}$Rb, $^{81}$Nb
and $^{85}$Tc are proton unstable, with upper limits of 100 ns and
less for their lifetimes. The stability of $^{77}$Y in this region is
particularly interesting and may be a consequence of the shape
polarizing effect of the $N= Z= 38$ core \cite{janas0}. With
increasing $Z$ one would expect these odd-$Z$ nuclei to become more
spherical and the odd proton to be more bound due to the influence of
the $N= Z= 50$ core. However, to our knowledge, $^{93}$Ag and
$^{97}$In have not been observed and $^{89}$Rh remains the heaviest
nucleus identified in this odd-$Z$ region so far.

Calculations of the one-proton separation energy $S_{\rm p}$ are
crucial for predicting the stability of isotopes near the proton drip
line. The $S_{\rm p}$ value tells about the relative stability of the
last occupied proton. The larger the value of $S_{\rm p}$, the more
proton stable is the nucleus. The nucleus is likely to be unstable
against proton emission if $S_{\rm p} < 0$. We have calculated the
one-proton separation energy from the ground-state binding energy of
two neighbouring nuclei using Eq.\ (\ref{eq1}) and show the results in
Table 5. We find that all of the nuclei considered here have a
positive $S_{\rm p}$ with the exception of $^{81}$Nb. One should note
that the determination of $S_{\rm p}$ arises from the difference of
two large numbers, and a small change in the ground-state energy may
alter the prediction. In this respect we should mention that the
effects of the pairing correlations for the even-even $(N,Z-1)$
systems used in Eq.\ (\ref{eq1}) to calculate $S_{\rm p}$ may be more
noticeable than in the pairing scheme adopted in our approximation.
Also, other corrections that we have not taken into account, such as
angular-momentum projection or correlations beyond mean field, like
fluctuations, may easily shift the value of $S_{\rm p}$ by several
hundred keV\@. Thus, in our calculation positive $S_{\rm p}$ values of
about or less than a half MeV can be considered compatible with having
a proton unstable system.

For all the parameter sets studied here, our RMF calculation
successfully predicts the stable nature of the nuclei $^{61}$Ga,
$^{65}$As, $^{77}$Y and $^{89}$Rh and the unstability of $^{81}$Nb
against proton emission. The conclusion is less definite in some cases
than in others (for example, $^{89}$Rh turns out to be stable in the
NL3 calculation but $S_{\rm p}$ is about only 0.5 MeV\@). The nuclei
$^{69}$Br, $^{73}$Rb and $^{85}$Tc are found to be stable ($S_{\rm p}
\sim 1$ MeV), contrary to the experimental evidence. The unobserved
$^{93}$Ag and $^{97}$In nuclei should be rather proton stable
according to the RMF calculations (maybe with some doubt in the case
of the TM1 set). The relativistic calculations indicate the stability
of the recently detected $^{79}$Zr and $^{83}$Mo isotopes, with the
only exception of $^{83}$Mo calculated with the NL-SH set. The
microscopic-macroscopic MMa calculations \cite{moller0} yield negative
$S_{\rm p}$ values for most of the odd-$Z$ nuclei of Table 5. The MMa
model predicts clearly that $^{79}$Zr and $^{83}$Mo are stable
systems, but fails to point out the stability of $^{77}$Y and
$^{89}$Rh. According to the MMa model $^{93}$Ag and $^{97}$In would be
proton unstable nuclei.

In Table 5 we also display the charge radius $r_{\rm ch}$ for the
ground-state solution. Taking into account the finite size of the
proton, it is obtained from the r.m.s.\ proton radius as $r_{\rm ch}
= \sqrt{r_{\rm p}^2 + 0.64}$ fm, where
\begin{equation}
 r_{\rm p}^2 = \frac{1}{Z}
 \int_0^\infty 2\pi r  dr \! \int_{-\infty}^\infty dz
\left[ r^2+z^2\right] \rho_{\rm p} (r,z)
\end{equation}
in cylindrical coordinates. For each nucleus the charge radii are
almost equal with all the four parameter sets (the changes are
generally less than 0.05 fm). We note that the magnitude of the
r.m.s.\ radii changes little between the solutions of different
deformation (again the changes are less than 0.05 fm, excluding
$^{77}$Y, $^{79}$Zr and $^{81}$Nb where we have found maximum
differences of $\sim 0.1$ fm between the various shapes). Hence we
only show the charge radii of the ground-state solutions.

\section{Summary and conclusions}
\hspace*{\parindent}
 
In summary, we have calculated the binding energy and the quadrupole
deformation parameter for odd $Z=N+1$, $T_z= -1/2$ nuclei in the
relativistic mean field model. The odd nucleon has been treated by the
blocking procedure. The spin of the intrinsic states of the blocked
nucleon, which is the resultant spin of the isotope, has been
determined. The RMF calculations produce two or three different
solutions for most of the $Z=N+1$ nuclei in the considered valley. In
some of the cases the isomeric solutions are very close to one another
and can be considered as coexistent shapes.

Shapes with large deformations are predicted near the proton drip line
in agreement with the microscopic-macroscopic calculations
\cite{janas0,moller0}. The spin of the $Z=N+1$ nuclei is well
reproduced in comparison with the microscopic-macroscopic model,
especially if one ignores the difference in binding energy between the
various shape-isomeric states. The one-proton separation energies are
found to be force dependent, but the four parameter sets studied
generally agree in the trends predicted for $S_{\rm p}$. Overall the
RMF predicts slightly bound configurations for the investigated
systems. In the case of the $^{81}$Nb nucleus $S_{\rm p}$ is negative
or zero, which indicates that the isotope is just beyond the stability
line. In the present calculations the nuclei $^{69}$Br, $^{73}$Rb and
$^{85}$Tc are proton bound. Experimentally these isotopes are
unstable, with estimated half-lives of less than about 100 ns
\cite{janas0}. The so far unobserved nuclei $^{93}$Ag and $^{97}$In
are found to be rather proton stable. We have checked for the NL3
parameter set that most of the nuclei studied in this work are the
predicted lightest stable odd isotopes. The exceptions are arsenic and
zirconium for which the lightest proton stable isotopes are $^{63}$As
and $^{77}$Zr, respectively.

We observe that the $S_{\rm p}$ values, and in some cases the energy
differences between oblate and prolate or spherical solutions, are of
the same order as the uncertainty in the binding energies of the
present RMF calculations. To further avoid ambiguities in the
prediction of separation energies and ground-state shapes, a more
sophisticated RMF approach for binding energy calculations would be
called for. In this connection the Dirac--Hartree--Bogoliubov approach
is a prescription to treat the pairing effects in a more proper way
\cite{meng0,meng1,lala1}. Finally, it should be remarked that in this
work we have been concerned with bulk properties, such as binding
energies, nuclear deformations and the average properties of the
intrinsic states, and not with the spectroscopy of the bands in the
studied nuclei. Therefore, only the intrinsic states have been
considered. To project out onto good angular momentum states remains
an interesting problem for future investigations of the relativistic
mean field model.

\bigskip \bigskip

The authors would like to acknowledge support from the DGICYT (Spain)
under grant PB98-1247 and from DGR (Catalonia) under grant
1998SGR-00011. S.K.P. thanks the Spanish Education Ministry grant
SB97-OL174874 for financial support and the Departament d'Estructura i
Constituents de la Mat\`eria of the University of Barcelona for kind
hospitality.

\newpage

\section*{Tables}

\begin{table}
\caption{RMF results for the binding energy ($B$), the quadrupole
deformation parameter ($\beta_2$), and the single-particle energy
$\epsilon_{\rm p}$ and Nilsson orbit $[Nn_{3}\Lambda]\Omega^{\pi}$ of
the state occupied by the odd proton are shown for the nuclei
$^{61}$Ga, $^{65}$As and $^{69}$Br. Results of microscopic-macroscopic
(MM) mass models are also given: MMa is from Ref.\ \cite{moller0} and
MMb is from Ref.\ \cite{janas0}. The energies are in MeV.}
\bigskip
\centering
\begin{tabular}{lcrcrcr}
\hline
\hline
&\multicolumn{1}{c}{set}
&\multicolumn{1}{c}{$\epsilon_{\rm p}$}
&\multicolumn{1}{c}{$[Nn_{3}\Lambda]\Omega^{\pi}$}
&\multicolumn{1}{c}{$\beta_2$}
&\multicolumn{1}{c}{$B$} \\
\hline

$^{61}$Ga&NL1 &$-$1.06&$[310]{1/2}^{-}$ &0.21 &513.7\\
& &$-$2.17&$[301]{1/2}^{-}$ &$-$0.13 &511.6\\
&NL-SH &$-$0.91&$[312]{3/2}^{-}$ &0.22 &513.1\\
& &$-$1.50&$[301]{3/2}^{-}$ &$-$0.18 &510.5\\
&TM1 &$-$1.61&$[312]{3/2}^{-}$ &0.23 &512.0\\
& &$-$2.16&$[301]{3/2}^{-}$ &$-$0.20 &509.9\\
&NL3 &$-$0.89&$[312]{3/2}^{-}$ &0.23 &511.3\\
& &$-$1.88&$[301]{3/2}^{-}$ &$-$0.19 &508.4\\
&MMa &&${1/2}^{-}$ &0.21 &\\

$^{65}$As&NL1 &$-$2.21&$[312]{3/2}^{-}$ &0.24 &542.7\\
& &$-$2.99&$[301]{1/2}^{-}$ &$-$0.25 &542.4\\
&NL-SH &$-$1.39&$[310]{1/2}^{-}$ &0.23 &541.9\\
& &$-$2.23&$[301]{1/2}^{-}$ &$-$0.23 &540.9\\
&TM1 &$-$1.95&$[310]{1/2}^{-}$ &0.24 &543.1\\
& &$-$2.78&$[301]{1/2}^{-}$ &$-$0.25 &542.5\\
&NL3 &$-$1.95&$[310]{1/2}^{-}$ &0.24 &540.2\\
& &$-$2.56&$[301]{1/2}^{-}$ &$-$0.24 &539.9\\
&MMa &&${3/2}^{-}$ &0.23 &\\

$^{69}$Br&NL1 &$-$0.94&$[404]{9/2}^{+}$ &$-$0.29 &575.5\\
& &$-$0.78&$[301]{3/2}^{-}$ &0.21 &574.2\\
&NL-SH &$-$1.52&$[404]{9/2}^{+}$ &$-$0.28 &573.1\\
& &$-$0.35&$[431]{1/2}^{+}$ &0.28 &571.6\\
&TM1 &$-$1.21&$[404]{9/2}^{+}$ &$-$0.29 &575.7\\
& &$-$0.44&$[303]{5/2}^{-}$ &0.22 &574.3\\
&NL3 &$-$1.23&[404]${9/2}^{+}$ &$-$0.29 &572.5\\
& &$-$0.20&$[431]{1/2}^{+}$ &0.28 &570.9\\
&MMa &&${9/2}^{+}$ &$-$0.32 &\\
&MMb &&$[404]{9/2}^{+}$ &$-$0.25 &\\

\hline
\hline
\end{tabular}
\end{table}

\begin{table}
\vspace*{-1.5cm}
\caption{Same as Table 1 for $^{73}$Rb, $^{77}$Y and $^{81}$Nb.}
\bigskip
\centering
\begin{tabular}{lcrcrcr}
\hline
\hline
&\multicolumn{1}{c}{set}
&\multicolumn{1}{c}{$\epsilon_{\rm p}$}
&\multicolumn{1}{c}{$[Nn_{3}\Lambda]\Omega^{\pi}$}
&\multicolumn{1}{c}{$\beta_2$}
&\multicolumn{1}{c}{$B$} \\
\hline

$^{73}$Rb&NL1  &$-$0.74&$[413]{7/2}^{+}$ &$-$0.35 &605.0\\
& &$-$2.23&$[431]{3/2}^{+}$ &0.42 &604.1\\
&NL-SH &$-$1.05&$[413]{7/2}^{+}$ &$-$0.34 &604.7\\
& &$-$2.34&$[431]{3/2}^{+}$ &0.41 &602.6\\
&TM1 &$-$0.73&$[404]{7/2}^{+}$ &$-$0.35 &605.8\\
& &$-$1.83&$[431]{3/2}^{+}$ &0.42 &603.4\\
&NL3 &$-$0.87&$[413]{7/2}^{+}$ &$-$0.35 &602.9\\
& &$-$2.16&$[431]{3/2}^{+}$ &0.42 &601.4\\
&MMa &&${3/2}^{+}$ &0.37 &\\
&MMb &&$[312]{3/2}^{-}$ &0.42 &\\

$^{77}$Y&NL1 &$-$1.01&$[422]{5/2}^{+}$ &0.49 &638.0\\
& &$-$1.35&$[330]{1/2}^{-}$ &$-$0.08 &637.7\\
&NL-SH &$-$1.22&$[422]{5/2}^{+}$ &0.47 &636.9\\
& &$-$1.15&$[404]{9/2}^{+}$ &$-$0.14 &631.4\\
& &$-$1.29&$p_{1/2}$ &0.00 &630.2\\
&TM1 &$-$0.67&$[404]{9/2}^{+}$ &$-$0.14 &636.9\\
& &$-$1.89&$p_{1/2}$ &0.00 &636.4\\
& &$-$0.67&$[422]{5/2}^{+}$ &0.49 &635.7\\
&NL3 &$-$1.00&$[422]{5/2}^{+}$ &0.48 &635.1\\
& &$-$0.84&$[404]{9/2}^{+}$ &$-$0.15 &632.6\\
& &$-$1.90&$p_{1/2}$ &0.00 &631.9\\
&MMa &&${5/2}^{+}$ &0.42 &\\
&MMb &&$[422]{5/2}^{+}$ &0.43 &\\

$^{81}$Nb&NL1 &0.04&$[404]{9/2}^{+}$ &$-$0.02 &670.6\\
& &$-$1.25&$[413]{7/2}^{+}$ &$-$0.21 &668.1\\
& &$-$0.42&[431]${1/2}^{+}$ &0.53 &667.8\\
&NL-SH &$-$0.02&$[431]{1/2}^{+}$ &0.52 &667.1\\
& &$-$1.63&$[413]{7/2}^{+}$ &$-$0.20 &663.8\\
& &$-$0.43&$[404]{9/2}^{+}$ &$-$0.02 &660.4\\
& TM1 &0.11&$[404]{9/2}^{+}$ &$-$0.02 &667.6\\
& &$-$1.11&$[413]{7/2}^{+}$ &$-$0.20 &667.3\\
& &$-$0.99&$[431]{1/2}^{+}$ &0.55 &664.9\\
& NL3 &$-$0.26&$[431]{1/2}^{+}$ &0.53 &664.7\\
& &$-$1.34&$[413]{7/2}^{+}$ &$-$0.20 &663.7\\
& &$-$0.07&$[404]{9/2}^{+}$ &$-$0.02 &663.5\\
&MMa &&${1/2}^{+}$ &0.46 &\\
&MMb &&$[431]{1/2}^{+}$ &0.44 &\\

\hline\hline
\end{tabular}
\end{table}

\begin{table}
\vspace*{-0.5cm}
\caption{Same as Table 1 for $^{85}$Tc, $^{89}$Rh, $^{93}$Ag and
         $^{97}$In.}
\bigskip
\centering
\begin{tabular}{lcrcrcr}
\hline\hline
&\multicolumn{1}{c}{set}
&\multicolumn{1}{c}{$\epsilon_{\rm p}$}
&\multicolumn{1}{c}{$[Nn_{3}\Lambda]\Omega^{\pi}$}
&\multicolumn{1}{c}{$\beta_2$}
&\multicolumn{1}{c}{$B$} \\
\hline

$^{85}$Tc&NL1 &$-$0.93&$[413]{5/2}^{+}$ &$-$0.22 &699.6\\
& &$-$1.27&$[431]{3/2}^{+}$ &0.09 &699.3\\
&NL-SH &$-$1.24&$[413]{5/2}^{+}$ &$-$0.22 &696.9\\
& &$-$0.40&$[301]{3/2}^{-}$ &0.31 &694.6\\
&TM1 &$-$0.71&$[413]{5/2}^{+}$ &$-$0.22 &698.1\\
& &$-$1.06&$[431]{3/2}^{+}$ &0.09 &696.0\\
&NL3 &$-$0.98&$[413]{5/2}^{+}$ &$-$0.22 &695.6\\
& &$-$1.35&$[431]{3/2}^{+}$ &0.09 &692.6\\
&MMa &&${3/2}^{+}$ &0.05 &\\
&MMb &&$[422]{5/2}^{+}$ &$-$0.25 &\\

$^{89}$Rh&NL1 &$-$1.30&$[422]{5/2}^{+}$ &0.16 &733.0\\
& &$-$0.46&$[411]{3/2}^{+}$ &$-$0.22 &730.1\\
&NL-SH &$-$0.62&$[411]{3/2}^{+}$ &$-$0.20 &728.7\\
& &$-$1.25&$[310]{1/2}^{-}$ &0.21 &726.2\\
&TM1 &$-$1.04&$[422]{5/2}^{+}$ &0.15 &728.4\\
& &$-$0.10&$[411]{3/2}^{+}$ &$-$0.21 &727.6\\
&NL3 &$-$1.33&$[422]{5/2}^{+}$ &0.15 &726.3\\
& &$-$0.40&$[411]{3/2}^{+}$ &$-$0.21 &726.2\\
&MMa &&${5/2}^{+}$ &0.05 &\\
&MMb &&$g_{9/2}$ &0.01 &\\

$^{93}$Ag&NL1 &$-$1.01&$[413]{7/2}^{+}$ &0.15 &766.4\\
& &$-$1.38&$[411]{3/2}^{+}$ &$-$0.08 &763.7\\
&NL-SH &$-$1.47&$[413]{7/2}^{+}$ &0.14 &760.6\\
& &$-$0.66&$[411]{1/2}^{+}$ &$-$0.18 &759.4\\
&TM1 &$-$0.82&$[413]{7/2}^{+}$ &0.14 &760.8\\
& &$-$1.14&$[411]{3/2}^{+}$ &$-$0.08 &758.5\\
&NL3 &$-$1.09&$[413]{7/2}^{+}$ &0.14 &759.9\\
& &$-$0.47&$[411]{1/2}^{+}$ &$-$0.18 &755.8\\
&MMa &&${7/2}^{+}$ &0.05 &\\

$^{97}$In&NL1 &$-$1.01&$[404]{9/2}^{+}$ &0.08 &800.2\\
&NL-SH &$-$1.46&$[404]{9/2}^{+}$ &0.07 &796.2\\
&TM1 &$-$0.84&$[404]{9/2}^{+}$ &0.08 &793.4\\
&NL3 &$-$1.10&$[404]{9/2}^{+}$ &0.08 &793.8\\
&MMa &&${9/2}^{+}$ &0.05 &\\

\hline
\hline
\end{tabular}
\end{table}

%
\begin{table}
\vspace*{-1.cm}
\caption{RMF results for the binding energy ($B$), the quadrupole
deformation parameter ($\beta_2$), and the single-particle energy
$\epsilon_{\rm n}$ and Nilsson orbit $[Nn_{3}\Lambda]\Omega^{\pi}$ of
the state occupied by the odd neutron are shown for the nuclei
$^{79}$Zr and $^{83}$Mo. The MMa values are from Ref.\ \cite{moller0}.
The energies are in MeV.}
\bigskip
\centering
\begin{tabular}{lcrcrcr}
\hline\hline
&\multicolumn{1}{c}{set}
&\multicolumn{1}{c}{$\epsilon_{\rm n}$}
&\multicolumn{1}{c}{$[Nn_{3}\Lambda]\Omega^{\pi}$}
&\multicolumn{1}{c}{$\beta_2$}
&\multicolumn{1}{c}{$B$} \\
\hline

$^{79}$Zr&NL1 &$-$15.57&$p_{1/2}$ &0.00 &655.0\\
& &$-$14.28&$[422]{5/2}^{+}$ &0.49&653.2\\
& &$-$14.19&$[404]{9/2}^{+}$ &$-$0.18 &652.3\\
&NL-SH  &$-$14.61&$[422]{5/2}^{+}$ &0.48 &652.6\\
& &$-$14.79&$[404]{9/2}^{+}$ &$-$0.18 &647.2\\
& &$-$14.31&$p_{1/2}$ &0.00 &645.7\\
&TM1 &$-$14.83&$p_{1/2}$ &0.00 &652.9\\
& &$-$14.20&$[404]{9/2}^{+}$ &$-$0.18 &651.8\\
& &$-$13.85&[422]${5/2}^{+}$ &0.50 &650.2\\
&NL3 &$-$14.31&$[422]{5/2}^{+}$ &0.49 &650.3\\
& &$-$14.92&$p_{1/2}$ &0.00 &648.5\\
& &$-$14.40&$[404]{9/2}^{+}$ &$-$0.18 &647.7\\
&MMa &&${5/2}^{+}$ &0.43 &\\

$^{83}$Mo&NL1 &$-$13.60&$[404]{9/2}^{+}$ &$-$0.05 &684.4\\
& &$-$14.61&[413]${7/2}^{+}$ &$-$0.22 &683.9\\
& &$-$15.64&$[301]{3/2}^{-}$ &0.27 &679.1\\
&NL-SH &$-$15.11&$[413]{7/2}^{+}$ &$-$0.21 &680.4\\
& &$-$14.26&$[303]{5/2}^{-}$ &0.38 &679.5\\
& &$-$14.26&$[404]{9/2}^{+}$ &$-$0.05 &675.2\\
&TM1 &$-$14.49&$[413]{7/2}^{+}$ &$-$0.22 &682.8\\
& &$-$13.58&$[404]{9/2}^{+}$ &$-$0.05 &681.3\\
& &$-$14.69&$[301]{3/2}^{-}$ &0.27 &678.4\\
&NL3 &$-$14.75&$[413]{7/2}^{+}$ &$-$0.22 &679.7\\
& &$-$13.79&$[404]{9/2}^{+}$ &$-$0.05 &677.6\\
& &$-$14.80&$[301]{3/2}^{-}$ &0.27 &675.5\\
&MMa &&${7/2}^{+}$ &$-$0.21 &\\
\hline
\hline
\end{tabular}
\end{table}

\newpage

\begin{table}
\caption{One-proton separation energies $S_{\rm p}$ (in MeV) and
charge radii $r_{\rm ch}$ (in fm) for the nuclei of Tables 1--4.
The MMa values are from Ref.\ \cite{moller0}. In the last column we
indicate whether there is experimental evidence of the proton
stability of the nucleus in question.} 
\bigskip
\centering
\begin{tabular}{lrcrcrcrcrc}
\hline
\hline
&\multicolumn{2}{c}{NL1}
&\multicolumn{2}{c}{NL-SH}
&\multicolumn{2}{c}{TM1}
&\multicolumn{2}{c}{NL3}
&\multicolumn{1}{c}{MMa}\\
&\multicolumn{1}{c}{$S_{\rm p}$}
&\multicolumn{1}{c}{$r_{\rm ch}$}
&\multicolumn{1}{c}{$S_{\rm p}$}
&\multicolumn{1}{c}{$r_{\rm ch}$}
&\multicolumn{1}{c}{$S_{\rm p}$}
&\multicolumn{1}{c}{$r_{\rm ch}$}
&\multicolumn{1}{c}{$S_{\rm p}$}
&\multicolumn{1}{c}{$r_{\rm ch}$}
&\multicolumn{1}{c}{$S_{\rm p}$}
&\multicolumn{1}{c}{Stable}\\
\hline

$^{61}$Ga& 0.92&3.90&0.77 &3.89&1.50 &3.92&0.77&3.90&$-$0.09
&Yes\\
$^{65}$As& 2.06&4.02&1.30 &4.00&1.86 &4.02&1.78&4.01&0.13
&Yes\\
$^{69}$Br& 0.87&4.14&1.40 &4.10&1.16 &4.12&1.13&4.12&0.09
&No\\
$^{73}$Rb& 0.82&4.23&$$1.12 &4.19&0.76 &4.22&0.90&4.21&$-$0.31
&No\\
$^{77}$Y & 1.05&4.32&1.19 &4.29&0.58 &4.23&1.02&4.30&$-$0.26
&Yes \\
$^{81}$Nb&$-$0.08&4.31&$-$0.28 &4.39 &$-$0.17&4.28&0.01&4.40 &$-$1.00 
&No \\
$^{85}$Tc& 1.05&4.37&1.29 &4.34&0.77 &4.36&1.05&4.35&$-$0.66
&No \\
$^{89}$Rh& 1.28&4.41&0.66 &4.38&0.96 &4.40&0.45&4.39&$-$0.50
&Yes \\
$^{93}$Ag& 0.95&4.44&1.26 &4.42&0.69 &4.44&1.00&4.43&$-$0.49
&No \\
$^{97}$In& 0.77&4.46&1.13 &4.45&0.49 &4.48&0.67&4.46&$-$0.34
&No \\
$^{79}$Zr& 3.27&4.28&2.11 &4.32&2.41 &4.25&1.96&4.33&2.36
&Yes \\
$^{83}$Mo& 1.20&4.34&0.19 &4.31&1.78 &4.33&1.63&4.33&1.26
&Yes \\
\hline
\hline
\end{tabular}
\end{table}

\end{document}